\documentclass[apj,iop]{emulateapj}
\submitted{accepted for publication in The Astrophysical Journal on 2012 October 9}
\usepackage{graphicx}
\usepackage[dvipdfm]{hyperref}

\newcommand {\eg} {{\it e.g.}}

\newcommand {\be} {\begin{equation}}
\newcommand {\ee} {\end{equation}}
\newcommand {\bea} {\begin{eqnarray}}
\newcommand {\eea} {\end{eqnarray}}

\begin{document}

\title{Herschel PACS and SPIRE observations of blazar PKS~1510$-$089:\\a case for two blazar zones}
\shorttitle{Herschel observations of PKS~1510$-$089}

\author{Krzysztof~Nalewajko\altaffilmark{1}, Marek~Sikora\altaffilmark{2}, Greg~M.~Madejski\altaffilmark{3}, Katrina~Exter\altaffilmark{4}, Anna~Szostek\altaffilmark{3,5}, Ryszard~Szczerba\altaffilmark{6}, Mark~R.~Kidger\altaffilmark{7} and Rosario Lorente\altaffilmark{7}}
\shortauthors{Nalewajko et~al.}

\altaffiltext{1}{University of Colorado, UCB 440, Boulder, CO 80309, USA; {\tt knalew@colorado.edu}}
\altaffiltext{2}{Nicolaus Copernicus Astronomical Center, Bartycka 18, 00-716 Warsaw, Poland}
\altaffiltext{3}{Kavli Institute for Particle Astrophysics and Cosmology, SLAC National Accelerator Laboratory, Stanford University, 2575 Sand Hill Road M/S 29, Menlo Park, CA 94025, USA}
\altaffiltext{4}{Instituut voor Sterrenkunde, KU Leuven, Celestijnenlaan 200 D, B-3001 Leuven, Belgium}
\altaffiltext{5}{Astronomical Observatory, Jagiellonian University, 30-244 Krak{\'o}w, Poland}
\altaffiltext{6}{Nicolaus Copernicus Astronomical Center, Rabia{\'n}ska 8, 87-100, Toru{\'n}, Poland}
\altaffiltext{7}{{\it Herschel} Science Centre, ESAC, P.O. Box 78, 28691 Villanueva de la Ca\~nada, Madrid, Spain}

\begin{abstract}
We present the results of observations of blazar PKS~1510$-$089 with the \emph{Herschel} Space Observatory PACS and SPIRE instruments, together with  multiwavelength data from  \emph{Fermi}/LAT, \emph{Swift}, SMARTS and SMA. The source was found in a quiet state, and its far-infrared spectrum is consistent with a power-law with a spectral index of $\alpha\simeq 0.7$. Our \emph{Herschel} observations were preceded by two `orphan' gamma-ray flares. The near-infrared data reveal the high-energy cut-off in the main synchrotron component, which cannot be associated with the main gamma-ray component in a one-zone leptonic model. This is because in such a model the luminosity ratio of the External-Compton and synchrotron components is tightly related to the frequency ratio of these components, and in this particular case an unrealistically high energy density of the external radiation would be implied. Therefore, we consider a well-constrained two-zone blazar model to interpret the entire dataset. In this framework, the observed infrared emission is associated with the synchrotron component produced in the hot-dust region at the supra-pc scale, while the gamma-ray emission is associated with the External-Compton component produced in the broad-line region at the sub-pc scale. In addition, the optical/UV emission is associated with the accretion disk thermal emission, with the accretion disk corona likely contributing to the X-ray emission.
\end{abstract}


\section{Introduction}

Of the many classes of Active Galactic Nuclei (AGN), blazars offer the most direct insight into the extreme plasma physics of powerful relativistic jets. The spectra of blazars span the entire range of the electromagnetic radiation accessible to  observational techniques and are routinely observed in the radio, millimeter, near-infrared (NIR), optical, UV, X-ray and gamma-ray bands. Even with this enormous observational scope, we still lack a consistent theoretical picture of the dissipation and radiative processes responsible for this mostly non-thermal and strongly variable emission.

The far-infrared (FIR) window to the Universe is rarely accessible owing to scarce availability of suitable observatories. \cite{1983ApJ...273...58C} combined data from the Kuiper Airborne Observatory at $107\;{\rm\mu m}$ and $240\;{\rm\mu m}$, and $400\;{\rm\mu m}$ data from the UKIRT telescope, with other NIR, mm and radio observations to construct the full infrared spectral energy distribution (SED) of blazar 3C~273. Their observations can be modeled remarkably well with a single synchrotron component. Many blazars were observed by the Infrared Astronomical Satellite (IRAS; \eg~\citealt{1988AJ.....95..307I}) between $12\;{\rm\mu m}$ and $100\;{\rm\mu m}$. The interpretation of their infrared spectra as synchrotron emission was strengthened by the detection of significant variability in these sources. \cite{1998ApJ...503L.109H} observed some blazars with the Infrared Space Observatory between $5\;{\rm\mu m}$ and $200\;{\rm\mu m}$. Three of the blazars had spectra consistent with a single synchrotron component, while in 3C~279 a thermal component was tentatively detected. \cite{2011ApJS..195...19O} observed another prominent blazar, 3C~454.3, with the \emph{Spitzer} Space Telescope, using all three instruments -- IRS, IRAC and MIPS. They found hints of  complex structure in the spectral range of MIPS ($24\;{\rm\mu m}$ -- $160\;{\rm\mu m}$), which they interpreted as possible evidence for two independent synchrotron components. Another interesting result involving the \emph{Spitzer} data was reported by \cite{2012ApJ...754..114H}. They detected a sharp spectral break in blazar 3C~279 in the MIPS spectral range, with a very hard spectral index of $\alpha=0.35\pm0.23$ ($F_\nu\propto\nu^{-\alpha}$) between $70\;{\rm\mu m}$ and $160\;{\rm\mu m}$. Combined with the overall spectral shape and multiwavelength variability characteristics, this finding was also interpreted in terms of two distinct synchrotron components.

The structure of the synchrotron spectral component in blazars is of great importance for understanding the physical structure of the so-called ``blazar zone'' in relativistic AGN jets. It became clear that more detailed FIR observations of blazars are needed. A great opportunity came with the launch of the \emph{Herschel} Space Observatory. In this work, we present photometric observations of another prominent blazar, PKS~1510$-$089, with two \emph{Herschel} instruments -- PACS and SPIRE. These results are combined with the publicly available multiwavelength data from \emph{Fermi}/LAT, \emph{Swift}, SMARTS and the Submillimeter Array (SMA). PKS~1510$-$089 was observed previously in the mid-IR (MIR) band with \emph{Spitzer} (IRS, IRAC and MIPS) by \cite{2011ApJ...732..116M}, who looked for signatures of thermal emission from the dusty torus but found the source spectrum to be consistent with a power-law. It is also a prominent gamma-ray source. In the spring of 2009, it showed a series of strong gamma-ray flares that were probed by \emph{Fermi}/LAT \citep{2010ApJ...721.1425A} and AGILE \citep{2011A&A...529A.145D}. During this time, it was also detected at very-high energies ($\sim 0.1$ -- $1\;{\rm TeV}$) by the H.E.S.S. observatory \citep{Wagner10}, as one of a handful of FSRQs known at these energies.

In Section \ref{sec_obs}, we report on our \emph{Herschel} observations and other multiwavelength data. In Section \ref{sec_res}, we present the observational results, in particular multiwavelength light curves and quasi-simultaneous SEDs. In Section \ref{sec_mod}, we present our model of the broad-band SED of PKS~1510$-$089. In Section \ref{sec_dis}, we discuss how our results compare to previous studies of PKS~1510$-$089. Our conclusions are given in Section \ref{sec_con}.

In this work, symbols with a numerical subscript should be read as a dimensionless number $X_{\rm n}=X/(10^n\,{\rm cgs\,units})$. We adopt a standard $\Lambda$CDM cosmology with $H_0=71\;{\rm km\,s^{-1}\,Mpc^{-1}}$, $\Omega_{\rm m}=0.27$ and $\Omega_{\rm \Lambda}=0.73$, in which the luminosity distance to PKS~1510$-$089 ($z=0.36$) is $d_{\rm L}=1.91\;{\rm Gpc}$.

\section{Observations}
\label{sec_obs}

\subsection{Herschel}
\label{sec_obs_herschel}

We observed PKS~1510$-$089 with the \emph{Herschel} Space Observatory \citep{2010A&A...518L...1P}, using the PACS \citep{2010A&A...518L...2P} and SPIRE \citep{2010A&A...518L...3G} instruments, in 5 epochs denoted as `H1' -- `H5' between 2011 Aug 1 (MJD 55774) and 2011 Sep 10 (MJD 55814) -- see Table \ref{Tab:obslog}. The PACS and SPIRE observations for each epoch took place no more than one day apart.

\subsubsection{Data Reduction}
\label{sec_obs_herschel_red}

The PACS observations\footnote{The PACS Observer's Manual is available at \url{http://herschel.esac.esa.int/Docs/PACS/html/pacs_om.html}.} are mini-scan maps taken in pairs with  scan and cross-scan positional angles of $70^\circ$ and $110^\circ$, respectively, and with ten scan legs, each of $3.5\arcmin$ length and $2\arcsec$ separation. For each epoch, the PACS observations were repeated to cover the blue+red and the green+red bands. The characteristic wavelengths for the red, green and blue bands are $160\;{\rm\mu m}$, $100\;{\rm\mu m}$ and $70\;{\rm\mu m}$, respectively. Both medium and fast scan speeds were used. The SPIRE observations\footnote{The SPIRE Observer's Manual is available at \url{http://herschel.esac.esa.int/Docs/SPIRE/html/spire_om.html}.} used the standard small-scan map method, and each observation returned fluxes in three bands: short (PSW; $250\;{\rm\mu m}$), medium (PMW; $350\;{\rm\mu m}$) and long (PLW; $500\;{\rm\mu m}$). The details are recorded in Table \ref{Tab:obslog}, including the date, observation ID, and (for PACS) the filter and scan speed.

The PACS observations were reduced using HIPE, a \emph{Herschel}-specific software package \citep{2010ASPC..434..139O}. We used the {\tt Track 9} pipeline starting from {\tt Level 0}, and with the calibration tree {\tt v32}. The pipeline tasks included crosstalk correction, non-linearity correction and second-level de-glitching ({\tt mapDeglitch} task with the {\tt timeordered} option on). The background was removed using the high-pass filter method, adopting filter widths of 15, 20 and 35 readouts for blue, green and red bands, respectively; source masking radius of $25\arcsec$ for all bands; and drop size ({\tt pixfrac}) of 1. The map pixel sizes are $1.1\arcsec$, $1.4\arcsec$ and $2.1\arcsec$ for the blue, green and red bands, respectively.

\begin{table}[t]
\caption{\emph{Herschel} observing log for PKS~1510$-$089}
\label{Tab:obslog}
\centering
\begin{tabular}{ll|ccccc|c}
\hline\hline
&& \multicolumn{5}{c}{PACS} & SPIRE \\
epoch & MJD & filter & \multicolumn{2}{c}{OID} & \multicolumn{2}{c}{speed} & OID \\
&&& s & xs & s & xs \\
\hline
H1 & 55774 & r+b & 24997 & 24998 & m & f & 24992 \\
& ...      & r+g & 25102 & 25103 & f & f \\
H2 & 55790 & r+b & 26661 & 26662 & m & f & 26659 \\
& ...      & r+g & 26709 & 26710 & f & f \\
H3 & 55794 & r+b & 27007 & 27008 & f & m & 27002 \\
& ...      & r+g & 27041 & 27042 & f & f \\
H4 & 55806 -- 55807 & r+b & 27805 & 27806 & f & m & 27048 \\
& ...        & r+g & 27833 & 27834 & f & f \\
H5 & 55813 -- 55814 & r+b & 28361 & 28362 & f & m & 28355 \\
& ...        & r+g & 28393 & 28394 & f & f \\
\hline
\end{tabular}
\tablecomments{Individual observations have unique identifiers (OID).  For each observational epoch (first column), we performed four scanning-mode observations with PACS (s -- scan, xs -- cross-scan), each with a combination of two filters out of three (r -- red, g -- green, b -- blue) and one of two scanning speeds (m -- medium, f -- fast); as well as one small-map observation with SPIRE.}
\end{table}

The SPIRE observations were also reduced using the HIPE, with the {\tt Track 9} pipeline starting from {\tt Level 0.5}, and with the calibration version {\tt spire\_cal\_1}. To remove the background, we used the {\tt destriping} task with standard parameter settings. The maps were made with pixel sizes of of $6\arcsec$, $10\arcsec$ and $14\arcsec$ for the PSW, PMW and PLW bands, respectively. Background sources were fitted and removed with the source extractor routine (removing only those that appeared at all epochs). The SPIRE maps were converted to the units of Jy/pix, to match the units of the PACS maps, using the recommended beam-to-pixel size conversion factors.

\begin{table*}[ht]
\caption{Photometric results of \emph{Herschel} PACS and SPIRE observations of PKS~1510$-$089}
\label{Tab:fluxes}
\centering
\begin{tabular}{l|lll|lll}
\hline\hline
&& PACS &&& SPIRE \\
& blue (70$\mu$) & green (100$\mu$m) & red (160$\mu$m) & PSW (250$\mu$m) & PMW (350$\mu$m) & PLW (500$\mu$m)  \\
\hline
H1 (MJD 55774) &    $0.52\pm0.01$ & $0.36\pm0.02$ & $0.29\pm0.02$ & $0.67\pm0.02$ &
$0.83\pm0.02$ & $0.99\pm0.01$  \\  
H2 (MJD 55790) &    $0.46\pm0.01$ & $0.33\pm0.02$ & $0.22\pm0.02$ & $0.61\pm0.02$ &
$0.79\pm0.02$ & $0.96\pm0.01$ \\
H3 (MJD 55794) &    $0.48\pm0.01$ & $0.35\pm0.02$ & $0.25\pm0.02$ & $0.64\pm0.02$ &
$0.84\pm0.02$ & $1.02\pm0.01$ \\
H4 (MJD 55806 -- 55807) & $0.39\pm0.01$ & $0.28\pm0.02$ & $0.20\pm0.02$ & $0.55\pm0.02$ &
$0.72\pm0.02$ & $0.92\pm0.01$ \\
H5 (MJD 55813 -- 55814) & $0.38\pm0.01$ & $0.31\pm0.02$ & $0.22\pm0.02$ & $0.57\pm0.02$ &
$0.74\pm0.02$ & $0.93\pm0.01$ \\ 
\hline
\end{tabular}
\tablecomments{The fluxes are given in units of Jy.}
\end{table*}

\subsubsection{Photometry}
\label{sec_obs_herschel_phot}

All maps were measured for photometric fluxes using the aperture photometry with recommended aperture sizes for the source and the sky, and published aperture corrections\footnote{``PACS instrument and calibration'' -- \url{http://herschel.esac.esa.int/twiki/bin/view/Public/PacsCalibrationWeb}; ``SPIRE instrument and calibration'' -- \url{http://herschel.esac.esa.int/twiki/bin/view/Public/SpireCalibrationWeb}.}. For SPIRE, we obtained two additional flux measurements by fitting the source {\tt on the map} (a standard HIPE task) and {\tt along the timeline}\footnote{We used a script provided by the SPIRE team: {\tt bendoSourceFit\_v9.py}.}. The final adopted flux value is the mean of the three measurements, with the differences between results for each epoch never exceeding $0.01\;{\rm Jy}$ ($<2\%$). The calibration uncertainties are reported to be $5\%$ for PACS and $7\%$ for SPIRE. No color corrections were applied to the measured fluxes, the SPIRE and PACS calibration assume a source spectral index of $\alpha=1$ ($F_\nu\propto\nu^{-\alpha}$). 

In all six bands, we checked whether PKS~1510$-$089 is a point source. We computed the radial flux profiles by measuring the flux in apertures of increasing radius, and compared them to the flux profiles produced from the PSF maps for each instrument (PACS: from FITS files provided on the instrument public page; SPIRE: from a calibration file). In all cases, the blazar was compatible with a point source.

For the PACS maps, we had a mixture of the fast and medium scan speeds: observations in the green band were taken with the fast scan speeds only, and those in the blue and red bands were taken with either of the two scan speeds. The available aperture corrections have been produced only for the medium (and slow) scan speeds, hence for the fast scan speeds there is an additional uncertainty of a few percent in the flux measurement (based on a comparison of their PSFs: PACS team communication). In addition, the signal-to-noise ratio is slightly worse on the fast scan speed maps. Therefore, we used only the medium scan speed map fluxes for the red and blue bands, and the average of the fast scan speed map fluxes for the green band. To check on the difference in photometry between scan speeds, we compared the results for fast and medium scan speed maps in the red and green: the difference is not greater than $0.02\;{\rm Jy}$ for both bands. This value is similar to the typical flux measurement errors (see Table \ref{Tab:fluxes}).

For all maps we measure the scatter in the background as the standard deviation between about 8 apertures placed in background regions, which currently is the best method of estimating the flux measurement uncertainty. Since the background is devoid of any obvious traces of the interstellar medium, the observing mode is the same for all the epochs, and the data are reduced in the same way, we report only one average flux error for each of the 6 bands covered by PACS and SPIRE. The final results of the PACS and SPIRE photometry of PKS~1510$-$089 are reported in Table \ref{Tab:fluxes}.

\subsection{Fermi/LAT}
\label{sec_obs_fermi}

The \emph{Fermi}/LAT telescope \citep{2009ApJ...697.1071A} for most of 2011 operated in the scanning mode, observing the entire sky frequently and fairly uniformly. We used the standard analysis software package {\tt Science Tools v9r27p1}, with the instrument response functions {\tt P7SOURCE\_V6} \citep{2012arXiv1206.1896F}, the Galactic diffuse emission model {\tt gal\_2yearp7v6\_v0} and the isotropic background model {\tt iso\_p7v6source}. Events of the {\tt SOURCE} class were extracted from the region of interest (ROI) of $10^\circ$ radius centered on the position of PKS~1510$-$089 ($\alpha=228.2^\circ$, $\delta=-9.1^\circ$). The background model included 17 sources from the \emph{Fermi}/LAT Second Source Catalog \citep{2012ApJS..199...31N} within $15^\circ$ from PKS~1510$-$089; their spectral models are power-laws with the photon index fixed to the catalog values, and for sources outside the ROI the normalizations were also fixed. In addition, our source model included TXS~1530$-$131, located $6^\circ$ from PKS~1510$-$089, which was in a flaring state \citep{2011ATel..3579}. The free parameters of the source model are: the normalizations of all point sources within ROI, as well as of the diffuse components, and the photon indices of PKS~1510$-$089 and TXS~1530$-$131. The source flux was calculated with the unbinned maximum likelihood method, following standard recommendations\footnote{\url{http://fermi.gsfc.nasa.gov/ssc/data/analysis/documentation/Cicerone/}} (zenith angle $<100^\circ$ and the {\tt gtmktime} filter {\tt (DATA\_QUAL==1) \&\& (LAT\_CONFIG==1) \&\& ABS(ROCK\_ANGLE)<52}). Measurements with the test statistic ${\rm TS}\ge 10$ \citep{1996ApJ...461..396M} and with the predicted number of gamma rays $N_{\rm pred}\ge 3$ are presented in figures as data points. For the SEDs, we also plot 2-$\sigma$ upper limits calculated with a method described in Section 4.4 of \cite{2010ApJS..188..405A}.

To calculate the medium-term light curve, we selected events registered between MJD~55740 and MJD~55830 of reconstructed energy between $E_{\rm min}=200\;{\rm MeV}$ and $E_{\rm max}=300\;{\rm GeV}$. The spectrum of PKS~1510$-$089 was modeled with a power-law with a free photon index. The light curve is presented with overlapping 3-day bins with a 1-day time step. The $\nu F_{\rm\nu}$ values are calculated from the fitted power-law model at photon energies of $200\;{\rm MeV}$ and $2\;{\rm GeV}$.

To calculate the long-term light curve, we selected events between MJD~54700 and MJD~55840, and modeled them in 6-day bins. We used the same energy range as before, and the $\nu F_{\rm\nu}$ values correspond to the photon energy of $2\;{\rm GeV}$.

To calculate the SEDs, we selected events registered over 3-day time intervals (MJD~55789 -- 55792 for the `H2' state, MJD~55766 -- 55769 for the `F2' state; see below) in overlapping energy bins of equal logarithmic width and uniform logarithmic spacing:
\be
E_{\rm min}\left(\frac{E_{\rm max}}{E_{\rm min}}\right)^{(i-k)/N}
\le E \le
E_{\rm min}\left(\frac{E_{\rm max}}{E_{\rm min}}\right)^{i/N}
\ee
with $E_{\rm min}=100\;{\rm MeV}$, $E_{\rm max}=100\;{\rm GeV}$, $k=3$, $N=18$ and $i\in\{k,...,N\}$. Within each bin, the spectrum of PKS~1510$-$089 was modeled with a power-law with a fixed photon index determined from power-law fits in the broad energy range $200\;{\rm MeV}\le E\le 300\;{\rm GeV}$: $\Gamma_{\rm H2} = 2.44$ and $\Gamma_{\rm F2} = 2.30$.

\begin{table*}
\centering
\caption{Results of \emph{Swift}/XRT observations of PKS 1510$-$089}
\label{t:pksXRT}
\begin{tabular}{c c c  c c c c c c}
\hline\hline
Obs ID & Date & Exposure & Counts & Counts & 0.3--10 keV Flux & $\Gamma_{\rm X}$ & Norm & Reduced\\
		& MJD	&	s	& src & bkg & $10^{-12}$ erg cm$^{-2}$ s$^{-1}$ & & $10^{-3}$ keV$^{-1}$ cm$^{-2}$ s$^{-1}$ &$\chi^2$\\
\hline
  31173075 & 55735.20 & 1301.29 &   140 &     4 & $   7.42^{+ 1.58}_{-1.00} $ & $ 1.68^{+ 0.17}_{-0.17}$ & $ 1.05^{+ 0.14}_{-0.13}$ &  0.49\\[0.05in]
  31173076 & 55742.08 & 1010.44 &   100 &     3 & $   7.34^{+ 2.82}_{-1.59} $ & $ 1.48^{+ 0.33}_{-0.31}$ & $ 0.86^{+ 0.17}_{-0.16}$ &  2.26\\[0.05in]
  31173077 & 55746.87 & 2828.23 &   391 &    38 & $   8.38^{+ 1.04}_{-0.88} $ & $ 1.40^{+ 0.09}_{-0.09}$ & $ 0.90^{+ 0.07}_{-0.07}$ &  0.91\\[0.05in]
  31173078 & 55749.32 & 3853.72 &   535 &     6 & $   8.16^{+ 0.82}_{-0.71} $ & $ 1.53^{+ 0.08}_{-0.08}$ & $ 1.01^{+ 0.06}_{-0.06}$ &  1.47\\[0.05in]
   31173079 & 55758.13 & 288.34 &    20 &     1 & -- & -- & --  & --\\
  31173080 & 55760.18 & 1913.07 &   284 &     3 & $  10.00^{+ 1.69}_{-1.40} $ & $ 1.52^{+ 0.13}_{-0.13}$ & $ 1.22^{+ 0.10}_{-0.11}$ &  0.41\\[0.05in]
  31173081 & 55762.48 & 1667.35 &   224 &     5 & $   9.20^{+ 1.80}_{-1.48} $ & $ 1.39^{+ 0.14}_{-0.13}$ & $ 0.98^{+ 0.10}_{-0.10}$ &  0.81\\[0.05in]
  31173082 & 55766.68 & 1995.81 &   350 &    16 & $  10.18^{+ 1.69}_{-1.20} $ & $ 1.52^{+ 0.11}_{-0.10}$ & $ 1.23^{+ 0.09}_{-0.09}$ &  1.73\\[0.05in]
  31173083 & 55768.47 & 2166.31 &   370 &     9 & $   9.68^{+ 1.22}_{-1.01} $ & $ 1.60^{+ 0.10}_{-0.10}$ & $ 1.27^{+ 0.09}_{-0.09}$ &  0.98\\[0.05in]
  31173084 & 55770.51 & 1709.98 &   305 &    13 & $  10.61^{+ 1.42}_{-1.19} $ & $ 1.54^{+ 0.11}_{-0.11}$ & $ 1.32^{+ 0.11}_{-0.11}$ &  0.80\\[0.05in]
  31173085 & 55772.22 & 1898.03 &   291 &     4 & $   9.99^{+ 1.45}_{-1.26} $ & $ 1.39^{+ 0.11}_{-0.11}$ & $ 1.05^{+ 0.10}_{-0.10}$ &  0.86\\[0.05in]
  31173086 & 55774.19 & 2023.39 &   330 &     5 & $  10.04^{+ 1.50}_{-1.09} $ & $ 1.58^{+ 0.11}_{-0.11}$ & $ 1.30^{+ 0.11}_{-0.11}$ &  0.78\\[0.05in]
   31173087 & 55776.33 & 388.63 &    40 &     0 & -- & -- & --  & --\\
  31173088 & 55778.14 & 1702.46 &   202 &     5 & $   7.31^{+ 1.12}_{-0.99} $ & $ 1.63^{+ 0.14}_{-0.14}$ & $ 0.99^{+ 0.10}_{-0.10}$ &  0.47\\[0.05in]
  31173089 & 55782.58 & 1664.85 &   283 &     4 & $  10.42^{+ 1.26}_{-1.17} $ & $ 1.65^{+ 0.12}_{-0.12}$ & $ 1.43^{+ 0.13}_{-0.13}$ &  0.70\\[0.05in]
  31173090 & 55784.67 & 1777.68 &   349 &     4 & $  12.88^{+ 1.80}_{-1.41} $ & $ 1.46^{+ 0.10}_{-0.10}$ & $ 1.47^{+ 0.12}_{-0.12}$ &  0.57\\[0.05in]
  31173091 & 55786.51 & 2401.99 &   407 &     6 & $  11.79^{+ 1.54}_{-1.34} $ & $ 1.30^{+ 0.09}_{-0.09}$ & $ 1.12^{+ 0.08}_{-0.08}$ &  0.90\\[0.05in]
  31173092 & 55788.89 & 2474.71 &   363 &     5 & $   9.93^{+ 1.41}_{-1.21} $ & $ 1.29^{+ 0.10}_{-0.10}$ & $ 0.93^{+ 0.08}_{-0.08}$ &  0.97\\[0.05in]
  31173093 & 55790.49 & 2045.96 &   346 &    27 & $  10.85^{+ 1.63}_{-1.22} $ & $ 1.39^{+ 0.11}_{-0.11}$ & $ 1.14^{+ 0.10}_{-0.10}$ &  1.03\\[0.05in]
  31173094 & 55793.66 & 1130.79 &   181 &     8 & $  12.38^{+ 3.42}_{-2.44} $ & $ 1.31^{+ 0.17}_{-0.17}$ & $ 1.19^{+ 0.14}_{-0.14}$ &  1.28\\[0.05in]
  31173095 & 55796.19 & 2048.46 &   350 &     6 & $  12.61^{+ 1.64}_{-1.57} $ & $ 1.27^{+ 0.10}_{-0.10}$ & $ 1.17^{+ 0.10}_{-0.10}$ &  0.98\\[0.05in]
  31173096 & 55799.65 & 2033.42 &   284 &     5 & $   9.88^{+ 1.68}_{-1.32} $ & $ 1.25^{+ 0.11}_{-0.11}$ & $ 0.89^{+ 0.08}_{-0.08}$ &  1.18\\[0.05in]
\hline
\end{tabular}
\tablecomments{The spectrum was fitted in the $0.3$ -- $10\;{\rm keV}$ band with a model {\tt wabs*powerlaw}, the hydrogen column density $N_{\rm H}$ was fixed at $7.9\times 10^{20}$ cm$^{-2}$, and the free photon index  $\Gamma_{\rm X}$. The reported flux values correspond to the de-absorbed spectrum. Errors on the normalization parameter and $\Gamma_{\rm X}$ are $1\sigma$. Flux errors correspond to 90\% confidence interval using xspec script {\tt fluxerror.tcl}. Background counts are scaled to the size of the source region using {\tt backscal} keyword. Observations with less than 75 counts were not fitted.}
\end{table*}

\begin{table*}
\centering
\caption{Results of \emph{Swift}/UVOT photometry for PKS~1510$-$089}
\label{t:pksUVOT}
\begin{tabular}{cc  cc  cc  cc }
\hline\hline
Obs ID & Date & V & B & U & W1 & M2 & W2 \\ 
\hline
00031173075 & 55735.19& $  4.65 \pm  0.38 $& $  6.69 \pm  0.30 $& $  6.99 \pm  0.30 $& $  6.01 \pm  0.32 $& $  7.92 \pm  0.37 $& $  6.71 \pm  0.27 $\\
00031173076 & 55742.08& $  5.03 \pm  0.44 $& $  7.63 \pm  0.36 $& $  8.32 \pm  0.36 $& $  7.21 \pm  0.39 $& $  9.77 \pm  0.45 $& $  9.47 \pm  0.37 $\\
00031173077 & 55746.78& $  6.24 \pm  0.49 $& $  7.80 \pm  0.23 $& $  8.49 \pm  0.25 $& $  7.38 \pm  0.29 $& $ 10.12 \pm  0.18 $& $  9.65 \pm  0.28 $\\
00031173078 & 55749.18& $  6.85 \pm  0.26 $& $  8.54 \pm  0.20 $& $  9.24 \pm  0.23 $& $  7.83 \pm  0.30 $& $ 10.63 \pm  0.25 $& $ 10.09 \pm  0.27 $\\
00031173079 & 55758.13& -- & -- & $  8.95 \pm  0.59 $& $  6.83 \pm  0.33 $& -- & -- \\
00031173080 & 55760.14& $  5.15 \pm  0.32 $& $  7.22 \pm  0.26 $& $  7.72 \pm  0.27 $& $  6.65 \pm  0.30 $& $  8.72 \pm  0.34 $& $  8.38 \pm  0.27 $\\
00031173081 & 55762.34& $  6.67 \pm  0.39 $& $  7.88 \pm  0.32 $& $  8.38 \pm  0.31 $& $  7.66 \pm  0.35 $& $ 10.15 \pm  0.37 $& $  9.50 \pm  0.31 $\\
00031173082 & 55766.64& $  7.71 \pm  0.38 $& $  9.05 \pm  0.28 $& $  9.55 \pm  0.30 $& $  7.88 \pm  0.34 $& $ 10.75 \pm  0.35 $& $  9.39 \pm  0.30 $\\
00031173083 & 55768.43& $  6.56 \pm  0.34 $& $  8.13 \pm  0.27 $& $  8.86 \pm  0.28 $& $  7.70 \pm  0.33 $& $  8.90 \pm  0.43 $& $  9.64 \pm  0.29 $\\
00031173084 & 55770.38& $  6.39 \pm  0.35 $& $  8.07 \pm  0.27 $& $  8.88 \pm  0.29 $& $  7.97 \pm  0.34 $& $ 10.41 \pm  0.33 $& $  8.98 \pm  0.29 $\\
00031173085 & 55772.13& $  5.28 \pm  0.34 $& $  6.99 \pm  0.27 $& $  7.78 \pm  0.28 $& $  6.44 \pm  0.30 $& $  9.66 \pm  0.34 $& $  8.76 \pm  0.29 $\\
00031173086 & 55774.12& $  7.58 \pm  0.37 $& $  8.92 \pm  0.29 $& $  9.02 \pm  0.29 $& $  7.81 \pm  0.34 $& $ 10.04 \pm  0.34 $& $  9.25 \pm  0.29 $\\
00031173087 & 55776.47& -- & $  6.83 \pm  0.38 $& $  8.10 \pm  0.38 $& $  6.61 \pm  0.34 $& -- & $ 11.67 \pm  1.36 $\\
00031173088 & 55778.06& $  5.33 \pm  0.38 $& $  6.84 \pm  0.30 $& $  7.11 \pm  0.28 $& $  6.19 \pm  0.31 $& $  9.24 \pm  0.32 $& $  8.54 \pm  0.30 $\\
00031173089 & 55782.48& $  5.91 \pm  0.40 $& $  7.83 \pm  0.31 $& $  8.26 \pm  0.31 $& $  7.11 \pm  0.34 $& $  9.84 \pm  0.37 $& $  9.44 \pm  0.32 $\\
00031173090 & 55784.57& $  7.36 \pm  0.62 $& $  9.49 \pm  0.28 $& $ 10.31 \pm  0.30 $& $  8.94 \pm  0.37 $& $ 12.30 \pm  0.57 $& $ 11.27 \pm  0.33 $\\
00031173091 & 55786.31& $  8.86 \pm  0.37 $& $ 10.79 \pm  0.28 $& $ 11.49 \pm  0.31 $& $ 10.44 \pm  0.41 $& $ 12.31 \pm  0.35 $& $ 11.35 \pm  0.33 $\\
00031173092 & 55788.78& $  5.95 \pm  0.32 $& $  6.89 \pm  0.24 $& $  7.86 \pm  0.25 $& $  6.96 \pm  0.30 $& $  9.47 \pm  0.29 $& $  9.41 \pm  0.28 $\\
00031173093 & 55790.19& $  6.75 \pm  0.35 $& $  8.34 \pm  0.27 $& $  8.90 \pm  0.28 $& $  7.55 \pm  0.32 $& $ 10.73 \pm  0.33 $& $ 10.22 \pm  0.31 $\\
00031173094 & 55793.66& $  3.78 \pm  0.40 $& $  6.88 \pm  0.33 $& $  7.72 \pm  0.32 $& $  6.96 \pm  0.36 $& $  8.80 \pm  0.54 $& $  9.07 \pm  0.33 $\\
00031173095 & 55796.06& $  5.43 \pm  0.35 $& $  6.77 \pm  0.30 $& $  7.77 \pm  0.28 $& $  6.76 \pm  0.31 $& $  9.08 \pm  0.33 $& $  8.74 \pm  0.28 $\\
00031173096 & 55799.61& $  4.57 \pm  0.33 $& $  6.19 \pm  0.26 $& $  7.37 \pm  0.26 $& $  6.07 \pm  0.28 $& $  8.85 \pm  0.33 $& $  8.41 \pm  0.27 $\\
\hline
\end{tabular}
\tablecomments{We report flux densities $\nu F_\nu$ corrected for extinction, in units of $10^{-12}\;{\rm erg\,s^{-1}\,cm^{-2}}$. Quoted errors are a sum of 1$\sigma$ statistical and a systematic error added in quadrature. Systematic errors are of the order of 5\%.}
\end{table*}

\subsection{Swift}
\label{sec_obs_swift}

\subsubsection{XRT Data Analysis}
\label{sec_obs_swift_xrt}

We analyzed the \emph{Swift}/XRT data following the recommendations given in the ``Data Reduction Guide v1.2''. We used the {\tt ftools} software package v6.11, the \emph{Swift} calibration files from November 2011, and the {\tt xspec} program v12.7.0. We started from Level 1 event files, and reduced the data using the {\tt xrtpipeline} script with default screening and filtering criteria. With {\tt xrtpipeline}, we created the exposure maps and used them to correct the {\tt arf} files for dead columns. We extracted the source and the background spectra, using the {\tt xselect} program v2.4b, from a circular region centered on the source and with a radius of $47\arcsec$. The background came from an annulus centered on the source and with the inner and outer radii of $80\arcsec$ and $135\arcsec$, respectively. To obtain the spectral parameters, we fitted observations which have more than 75 source counts with an absorbed power-law model with hydrogen column density fixed at its Galactic value of $7.9\times 10^{20}\;{\rm cm}^{-2}$ \citep{2008ApJ...672..787K}, and with a free photon index. To calculate the flux errors, we used a script {\tt fluxerror.tcl} provided by the {\tt xspec} team\footnote{\url{http://heasarc.nasa.gov/xanadu/xspec/fluxerror.html}}. The results are given in Table \ref{t:pksXRT}.

\subsubsection{UVOT Data analysis}
\label{sec_obs_swift_uvot}

To analyze the UVOT data in the image mode, we followed the recommendations from the ``UVOT Software Guide v2.2'', and started from the Level 1 raw data. We constructed  a bad pixel map for each exposure to remove the bad pixels from further analysis. We reduced a {\tt modulo8} fixed-pattern noise from the images and the pixel-to-pixel fluctuations in the images due to detector sensitivity variations. Then, we converted the images from a raw coordinate system to tangential projection on the sky. Before adding the images, we applied an aspect-ratio correction to each exposure to obtain the correct sky coordinates of the UVOT sources and to ensure that individual exposures were added without offsets. Finally, we added all image exposures for a specific filter in a given observation. To extract the source magnitude and counts, we used the {\tt uvotsource} task. We used an aperture of $5\arcsec$ for all filters, which matches the aperture used to calibrate the UVOT photometry, and therefore it does not require aperture corrections. The background is estimated from an annulus with radii of $15\arcsec$ and $25\arcsec$ centered on the source. The results are presented in Table \ref{t:pksUVOT}.

To convert the observed magnitudes $m_\lambda$ to flux densities $\nu F_\nu$, we introduce an effective zero point $Z_\lambda$, such that $\nu F_\nu[{\rm erg\,s^{-1}\,cm^{-2}}]=10^{(Z_\lambda-m_\lambda+A_\lambda)/2.5}$, where $A_\lambda$ is the extinction. For \emph{Swift}/UVOT, we took
\bea
Z_\lambda^{\rm(UVOT)} &=& Z_\lambda^{\rm(P08)}+2.5\log(\lambda_{\rm eff}^{\rm(P08)}[{\rm \AA}]\times C_{\rm F}^{\rm(P08)}
\nonumber\\&&
[{\rm erg\,s^{-1}\,cm^{-2}\,\AA^{-1}}])\,,
\eea
where $Z_\lambda^{\rm(P08)}$, $\lambda_{\rm eff}^{\rm(P08)}$ and $C_{\rm F}^{\rm(P08)}$ are parameters taken from Tables 6, 8 and 10 in \cite{2008MNRAS.383..627P}, respectively. We adopt $\lambda_{\rm eff}^{\rm(P08)}$ as the effective wavelength for the UVOT filters. For the extinction correction in the direction to PKS~1510$-$089, we adopted a standard Galactic extinction model by \cite{cardelli1989} with parameters $E(B-V)=0.101$ and $R_{\rm V}=3.1$. In Table \ref{t:factorspks}, we report the effective wavelengths, extinctions and effective zero points for UVOT filters U, W1, M2 and W2. The effective zero points for filters V and B are consistent with the Cousins-Glass-Johnson photometric system discussed in Section \ref{sec_obs_smarts_sma}.

\begin{table}
\centering
\caption{Effective wavelengths, extinctions and effective zero points for the SMARTS (K --- B) and \emph{Swift}/UVOT (U --- W2) filters}
\label{t:factorspks}
\begin{tabular}{cccc}
\hline\hline
Filter & $\lambda_{\rm eff}$ [${\rm\mu m}$] & $A_{\lambda}$ [mag] & $Z_{\rm\lambda,eff}$ \\
\hline
K  & 2.19  & 0.04 & -15.14 \\
J  & 1.22  & 0.09 & -13.52 \\
R  & 0.641 & 0.26 & -12.11 \\
V  & 0.545 & 0.32 & -11.75 \\
B  & 0.438 & 0.42 & -11.39 \\
\hline
U  & 0.350 & 0.50 & -12.27 \\
W1 & 0.263 & 0.66 & -12.45 \\
M2 & 0.223 & 0.96 & -12.49 \\
W2 & 0.203 & 0.92 & -12.40 \\
\hline
\end{tabular}
\end{table}

\subsection{SMARTS and SMA}
\label{sec_obs_smarts_sma}

We used public optical and NIR data (B, V, R, J and K filters) from the Yale University SMARTS project\footnote{\url{http://www.astro.yale.edu/smarts/glast/}}. A part of the data for PKS~1510$-$089 was presented in \cite{2012ApJ...756...13B}. The magnitudes $m_{\rm\lambda}$ were converted into flux densities using the effective zero points introduced in Section \ref{sec_obs_swift_uvot}, here calculated as
\bea
Z_\lambda^{\rm(SMARTS)} &=& 2.5\log(\nu_{\rm eff}[{\rm Hz}]\times f_{\rm\nu}^{\rm(B98)}
\nonumber\\&&
[{\rm erg\,s^{-1}\,cm^{-2}\,Hz^{-1}}])\,,
\eea
where $\nu_{\rm eff}=c/\lambda_{\rm eff}^{\rm(B98)}$, and $\lambda_{\rm eff}^{\rm(B98)}$ and $f_{\rm\nu}^{\rm(B98)}$ are parameters of the Cousins-Glass-Johnson photometric system taken from Table A2 in \cite{1998A&A...333..231B}. The effective wavelengths and zero points for each SMARTS filter are reported in Table \ref{t:factorspks}.

We obtained the SMA data for PKS~1510$-$089 at $1.3\,{\rm mm}$ wavelength from the SMA Callibrator List\footnote{\url{http://sma1.sma.hawaii.edu/callist/callist.html}} \citep{2007ASPC..375..234G}. We use these data only to verify that they lie on the power-law extrapolation of the \emph{Herschel} PACS and SPIRE SED.

\section{Results}
\label{sec_res}

\subsection{Herschel PACS and SPIRE}

\begin{figure}
\includegraphics[width=\columnwidth]{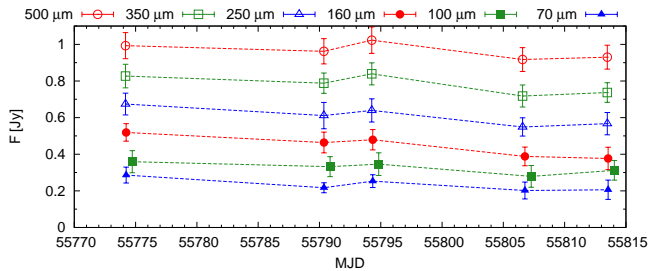}
\caption{\emph{Herschel} PACS (\emph{filled symbols}) and SPIRE (\emph{empty symbols}) light curves of PKS~1510$-$089.}
\label{fig_lc_pacs_spire}
\end{figure}

Figure \ref{fig_lc_pacs_spire} shows the light curves of PKS~1510$-$089 calculated for each filter of the PACS and SPIRE instruments. The source was not significantly variable over weekly time scales across the entire spectral range. The slight variations observed at different wavelengths appear to be correlated.

\begin{figure}
\includegraphics[width=\columnwidth]{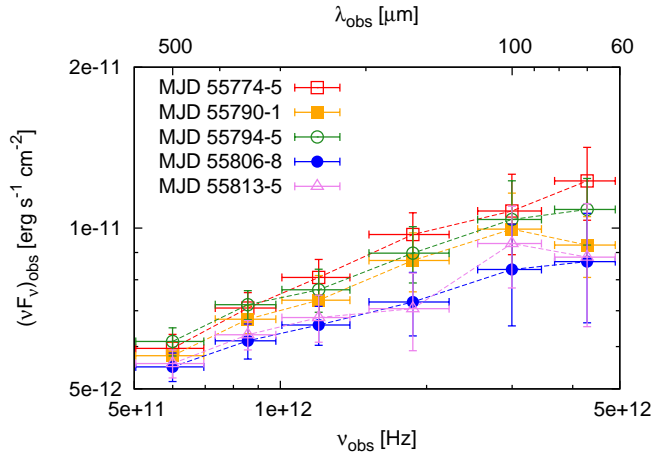}
\caption{\emph{Herschel} PACS and SPIRE quasi-simultaneous spectral energy distributions of PKS~1510$-$089.}
\label{fig_sed_pacs_spire}
\end{figure}

Figure \ref{fig_sed_pacs_spire} shows the FIR spectral energy distributions (SEDs) of PKS~1510$-$089 in 5 epochs; for each epoch observations in all six bands were performed within one day. These SEDs are generally consistent with power laws. The parameters of the spectral fits for each observational epoch are reported in Table \ref{tab_sed_fits}. A slight harder-when-brighter trend is apparent, although the range of parameter values is rather small.

\begin{table}
\centering
\caption{Spectral fits to the \emph{Herschel} PACS and SPIRE data}
\label{tab_sed_fits}
\begin{tabular}{ccc}
\hline\hline
epoch     & $\alpha$        & $\log(\nu F_\nu\;[{\rm erg\,s^{-1}\,cm^{-2}}])$ at $10^{12}\;{\rm Hz}$ \\
\hline
H1 & $0.64  \pm 0.02$  & $-11.130 \pm 0.007$ \\
H2 & $0.74  \pm 0.04$  & $-11.158 \pm 0.019$ \\
H3 & $0.71  \pm 0.02$  & $-11.135 \pm 0.007$ \\
H4 & $0.766 \pm 0.012$ & $-11.201 \pm 0.004$ \\
H5 & $0.75  \pm 0.04$  & $-11.192 \pm 0.014$ \\
\hline
\end{tabular}
\tablecomments{The model is a power-law function in the form $F_\nu\propto \nu^{-\alpha}$.}
\end{table}

\begin{figure*}
\centering
\includegraphics[width=0.85\textwidth]{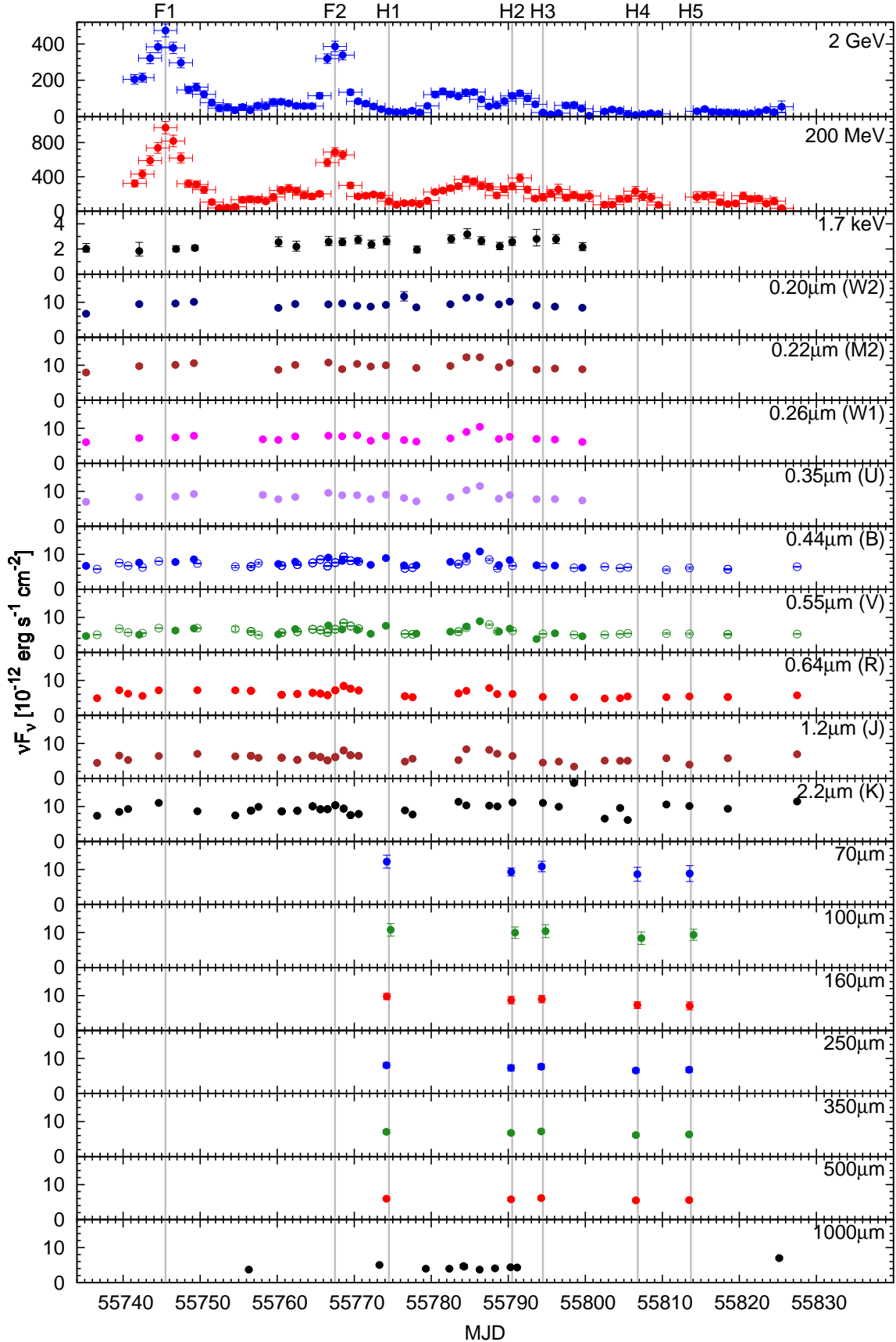}
\caption{Multiwavelength light curves of PKS~1510$-$089, including data from SMA, \emph{Herschel} PACS and SPIRE, SMARTS, \emph{Swift} UVOT and XRT, and \emph{Fermi}/LAT. The \emph{Swift}/XRT and \emph{Fermi}/LAT light curves are calculated at the indicated photon energies from power-law fits over broader energy ranges ($0.3$ --- $10\;{\rm keV}$ and $0.2$ --- $300\;{\rm GeV}$, respectively). All panels use the same units. The SMARTS and \emph{Swift}/UVOT data are corrected for extinction. \emph{Vertical lines} indicate the two \emph{Fermi} flares (`Fn') and the five \emph{Herschel} epochs (`Hn').}
\label{fig_lc_mwl}
\end{figure*}

\begin{figure}
\centering
\includegraphics[width=\columnwidth]{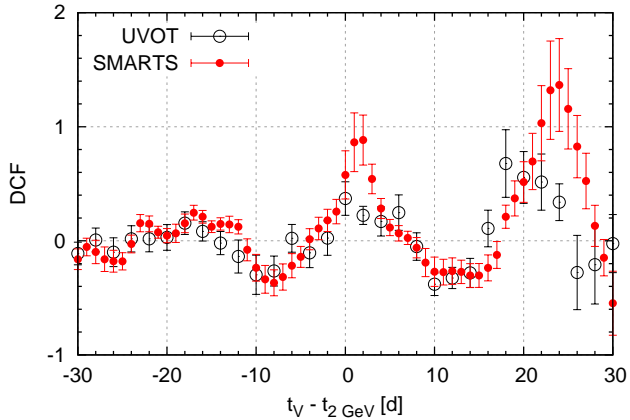}
\caption{Discrete correlation function (DCF; \citealt{1988ApJ...333..646E}) calculated between the \emph{Fermi}/LAT light curve at $2\;{\rm GeV}$ and the optical light curves in the $V$ band using either SMARTS ($\Delta t=1\;{\rm d}$) or \emph{Swift}/UVOT ($\Delta t=2\;{\rm d}$) data (see Figure \ref{fig_lc_mwl}). Positive argument values indicate optical signal delayed with respect to the gamma-ray signal.}
\label{fig_dcf}
\end{figure}

\begin{figure*}
\includegraphics[width=\textwidth]{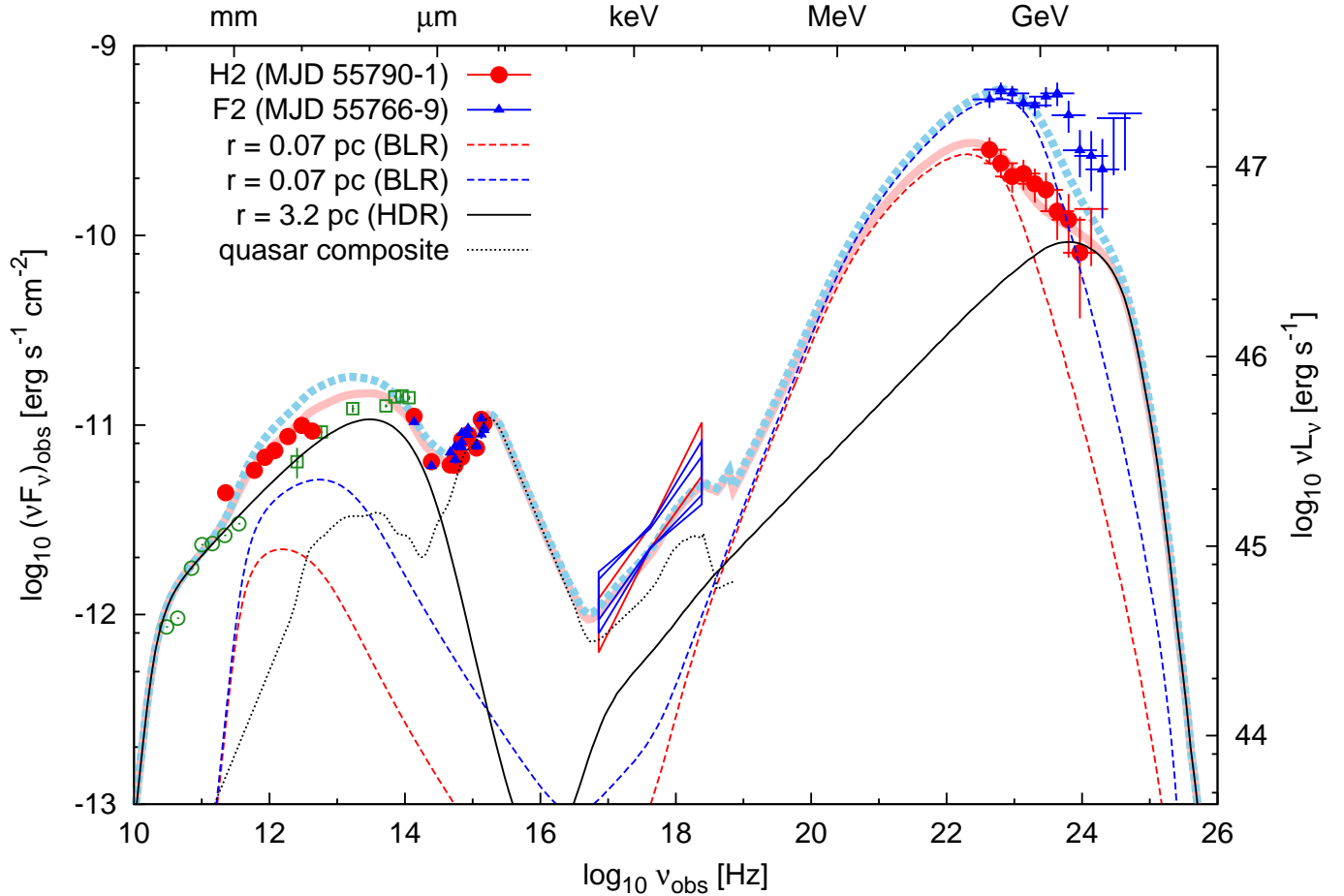}
\caption{Broad-band spectral energy distribution of PKS~1510$-$089 extracted for the second Herschel epoch (H2; \emph{red filled circles}) and for the second gamma-ray flare (F2; \emph{blue triangles}). The SMARTS and \emph{Swift}/UVOT data are corrected for extinction, using values from Table \ref{t:factorspks}. The \emph{green open circles} show the \emph{Planck} spectrum registered in 2010 Feb \citep{2011A&A...536A..15P}, and the \emph{green open squares} show the \emph{Spitzer} spectrum registered in 2007 Aug-Sep \citep{2011ApJ...732..116M}. We also plot our SED model for the H2 state (\emph{thick solid light-red ribbon}) and for the F2 state (\emph{thick dashed light-blue ribbon}). Their three components are: a slowly variable component produced beyond the edge of the dusty torus (\emph{solid lines}; synchrotron radiation for $\nu_{\rm obs,16}<1$; EC(HDR) radiation otherwise; negligible contribution from SSC and EC(BLR)), a rapidly variable component produced in the broad-line region (\emph{dashed lines} --- \emph{red} for H2; \emph{blue} for F2; synchrotron radiation for $\nu_{\rm obs,16}\lesssim 1$; EC(BLR) otherwise; marginal contribution from SSC at $\nu_{\rm obs,17}\sim 1$ for F2; negligible contribution from EC(HDR)), and a composite spectrum of radio-loud quasars (\citealt{1994ApJS...95....1E}, \emph{dotted line}).}
\label{fig_sed}
\end{figure*}

\begin{figure}
\includegraphics[width=\columnwidth]{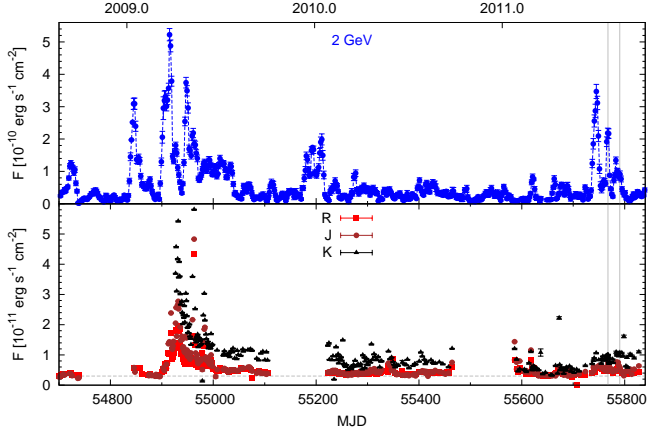}
\caption{Long-term light curves of PKS~1510$-$089. The \emph{top panel} shows the \emph{Fermi}/LAT flux modeled in 6-day time intervals by fitting a power-law in the energy range between $200\;{\rm MeV}$ and $300\;{\rm GeV}$, and taking the $\nu F_\nu$ value corresponding to $2\;{\rm GeV}$. The \emph{bottom panel} shows the optical/NIR data from SMARTS. The \emph{vertical lines} mark epochs F2 and H2, for which the broad-band SEDs were extracted.}
\label{fig_lc_long}
\end{figure}

\begin{figure}
\includegraphics[width=\columnwidth]{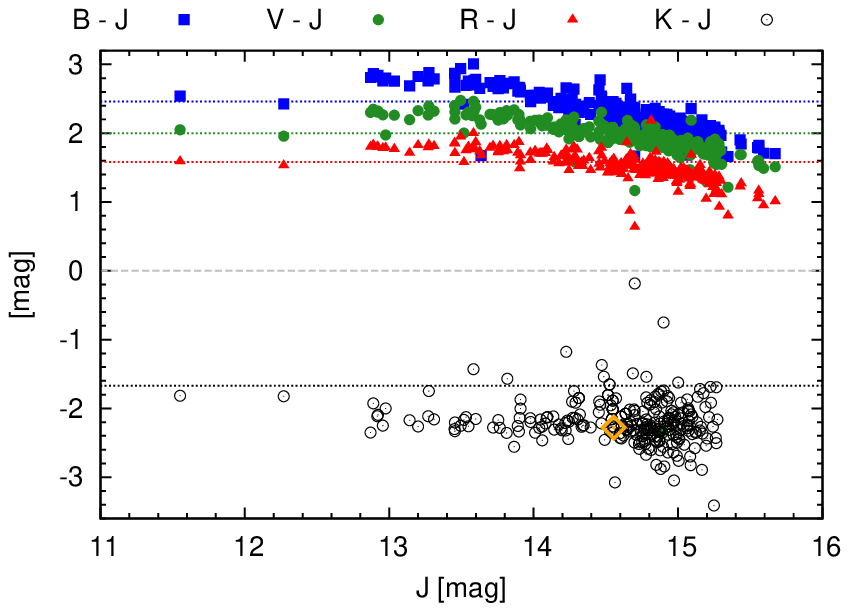}
\caption{Optical/NIR colors of PKS~1510$-$089 plotted against the $J$ magnitude based on the long-term SMARTS data. The \emph{dotted lines} mark the color values corresponding to a flat SED ($\nu F_\nu$). The \emph{orange point} marks the $K-J$ color corresponding to the second \emph{Herschel} epoch (H2; MJD~55790).}
\label{fig_colors}
\end{figure}

\subsection{Multiwavelength data}

To provide a context for the results obtained with \emph{Herschel}, we analyze quasi-simultaneous multiwavelength data for PKS~1510$-$089: gamma-ray data from \emph{Fermi}/LAT, optical/UV and X-ray data from \emph{Swift}, optical/NIR data from SMARTS, and millimeter data from SMA. In Figure \ref{fig_lc_mwl}, we present multiwavelength light curves calculated over a period of $\sim 3$ months encompassing our \emph{Herschel} campaign. The simultaneous multiwavelength coverage varies between different \emph{Herschel} pointings. The \emph{Herschel} observations span a period of relatively low activity following two prominent gamma-ray flares -- `F1' (MJD~55745; \citealt{2011ATel..3473}) and `F2' (MJD~55767).\footnote{Two more prominent gamma-ray flares were observed in PKS~1510$-$089 in Oct--Nov 2011 (Orienti et~al., submitted).} The \emph{Fermi}/LAT data indicate a modest spectral variability across the gamma-ray band. The F2 gamma-ray flare has a possible optical/NIR counterpart seen in the SMARTS data \citep[see also][]{2011ATel..3509}. The discrete correlation function \citep{1988ApJ...333..646E} calculated between the \emph{Fermi}/LAT data at $2\;{\rm GeV}$ and the SMARTS data in the $V$ band (Figure \ref{fig_dcf}) indicates that the optical flux is delayed with respect to the gamma-ray flux by $1-2$ days. However, such correlation is not confirmed by the \emph{Swift}/UVOT data, the brighter F1 gamma-ray flare does not have a similar optical counterpart, and the amplitude of optical variability is one order of magnitude smaller than the amplitude of the gamma-ray flare. Thus, these two gamma-ray flares can be practically called `orphan' flares. Of the NIR/optical/UV bands, the most prominent activity is seen in the K band.

We extracted the broad-band spectral energy distributions (SEDs) of PKS~1510$-$089 for two epochs. The second \emph{Herschel} pointing (H2) is chosen among other \emph{Herschel} pointings for the best overall multiwavelength coverage and the highest simultaneous gamma-ray flux. The second gamma-ray flare (F2) has a better multiwavelength coverage than the first gamma-ray flare. These two SEDs are shown in Figure \ref{fig_sed}. We find a very good agreement in the NIR/optical/UV and X-ray bands between these two epochs. There is a prominent difference in the gamma-ray band, not only in the integrated luminosity, but also in the spectral shape. In the low gamma-ray state (H2), the gamma-ray spectrum is much softer and can be reasonably approximated with a single power law. In the high gamma-ray state (F2), a possible double structure is seen, with peaks at $\sim 250\;{\rm MeV}$ and $\sim 1.5\;{\rm GeV}$, and a dip at $\sim 700\;{\rm MeV}$.\footnote{Whether this is a real spectral feature or just a statistical fluctuation, requires a more detailed analysis. Our conclusions do not rely on this issue.} The spectrum in the F2 state is significantly harder, at least up to $\sim 3\;{\rm GeV}$. Because of such a hard spectrum, the integrated luminosity calculated by fitting a power-law model up to $100\;{\rm GeV}$ might be significantly overestimated.

The FIR/mm spectrum probed by the \emph{Herschel} and SMA is consistent with a simple power law. While the highest-frequency PACS point ($70\;{\rm\mu m}$) in the H2 state indicates a small discrepancy from this trend, the \emph{Herschel} data at other epochs do not show any persistent spectral structure there. We note that the spectral index measured by \emph{Herschel} is consistent with the non-simultaneous observations in overlapping spectral windows by \emph{Planck} and \emph{Spitzer}. Such a well-aligned power-law spectrum can be naturally explained with a single synchrotron component in the optically thin regime. An interesting question is how this component connects to the NIR/optical data. In the NIR band, the SMARTS data indicate a peculiarly soft spectrum between $K$ and $J$ bands, as compared to a hard optical/UV spectrum between $R$ and $W2$ bands (a similar NIR spectrum can be seen in the data presented by \citealt{1988AJ.....95..307I}). Such feature can be understood only as the high-energy end of the synchrotron component. In Section \ref{sec_mod}, we consider a model in which the FIR and NIR spectra are connected with a single synchrotron component.

\subsection{Long-term variability}

In Figure \ref{fig_lc_long}, we compare the long-term light curves collected in the gamma-ray and optical/NIR bands by  \emph{Fermi}/LAT and SMARTS, respectively. These data include a previous active period in the first half of 2009 analyzed in detail by \cite{2010ApJ...710L.126M}, \cite{2010ApJ...721.1425A} and \cite{2011A&A...529A.145D}, and they partially overlap with the optical/NIR data from SMARTS analyzed by \cite{2012ApJ...756...13B} and \cite{2012ApJ...749..191C}. In 2009, a series of gamma-ray flares was accompanied by prominent optical/NIR activity, in contrast to the situation observed in 2011.

The long-term SMARTS data indicate the existence of a lower limit to the optical flux at the level of $F_{\rm min}\simeq 3\times 10^{-12}\;{\rm erg\,s^{-1}\,cm^{-2}}$ \citep[see also][]{2010ApJ...710L.126M}. In the $J$ and $R$ bands (but also in $V$ and $B$), this flux level was significantly exceeded only in the 2009 active state. The long-term constancy of the optical flux in quiet states indicates that it is not associated with the relativistic jet, but rather it is dominated by the thermal emission of the accretion disk. On the other hand, in the 2009 flaring state, the optical flux is most likely associated with  jet synchrotron emission. The lack of correlated optical activity corresponding to the gamma-ray flares in the summer of 2011 can be explained by a low level of the synchrotron component in the optical/NIR band. We will use these clues in our attempt to model the broad-band SED.

The long-term light-curve in the $K$ band shows a somewhat distinct behavior from the $J$ band and higher frequencies. The $K$ flux approaches $F_{\rm min}$ only in early 2011, and shows stronger and faster variability in the quiet state. In Figure \ref{fig_colors}, we present a color-luminosity diagram based on the whole SMARTS dataset for PKS~1510$-$089. We find that, while the $B-J$, $V-J$ and $R-J$ colors have a clear trend of being ``redder-when-brighter'', the $K-J$ color shows no such behavior. The $K$ --- $J$ part of the $\nu F_\nu$ SED is consistently soft, while the $J$ --- $B$ part is soft at high luminosities and hard at low luminosities. It appears that in the quiet state the $K$ luminosity is rather poorly correlated with other SMARTS bands. All the above evidence suggests that the $K$ band marks the high-energy cut-off/break of the synchrotron component.

\section{Modeling the broad-band SED}
\label{sec_mod}

In this Section, we attempt to model the broad-band SED of PKS~1510$-$089 during our second \emph{Herschel} epoch (H2), as presented in Figure \ref{fig_sed}. We employ the leptonic radiative code {\tt Blazar} \citep{2003A&A...406..855M}, which incorporates the exact treatment of the inverse-Compton emission in the Klein-Nishina regime, synchrotron self-absorption and pair-production absorption. {\tt Blazar} calculates the evolution of electrons injected at a constant rate over a distance range between $r_0/2$ and $r_0$ into a relativistically propagating spherical shell. The resulting non-thermal radiation is integrated over the same scale, and effectively it is dominated by the contribution from $r_0$. The variability properties of the source, with gamma-ray flares having no corresponding activity in the optical and IR bands, indicate that more than one emitting region is present. However, we begin by considering a one-zone model and a formal discussion of the physical constraints imposed by it.

In the optical/UV band, the hard spectrum, the lack of long-term flux variations, and the presence of a lower limit on the observed flux favor the dominance of a thermal component. In Figure \ref{fig_sed}, we plot the composite spectrum for radio-loud quasars from \cite{1994ApJS...95....1E}, normalizing it to the observed UV flux. We note that the composite spectrum nicely matches the observed optical/UV spectral index of PKS~1510$-$089, and it is also in reasonable agreement with the simultaneous X-ray spectrum. Although the observed X-ray flux is higher than the normalized composite spectrum by factor $\sim 1.6$, taking into account all the uncertainties and caveats involved in calculating the composite spectrum -- in particular the observed scatter of the UV/X-ray luminosity ratio in quasars -- we consider this discrepancy to be marginal. Therefore, at least a partial contribution of the hot accretion disk corona to the observed X-ray emission is likely, and this can explain the relatively low variability amplitude observed in PKS~1510$-$089 in the X-ray band over several years \citep{2010ApJ...710L.126M}.

The bolometric luminosity of the accretion disk is estimated by integrating the normalized spectrum of the quasar composite, excluding its infrared and X-ray components, which yields $L_{\rm d}\simeq 5\times 10^{45}\;{\rm erg\,s^{-1}}$. Using this value, we can estimate the characteristic radii of the broad-line region (BLR): $r_{\rm BLR}\simeq 0.07\;{\rm pc}$; and the hot-dust region (HDR): $r_{\rm HDR}\simeq 2.9\;{\rm pc}\times T_3^{-2.6}$ \citep[see][]{2008MNRAS.386..945T,2008ApJ...685..160N,2009ApJ...704...38S}. Within these radii, the energy density of external radiation fields is roughly independent of the radius, and in the external frame is given by $u_{\rm ext}\simeq \xi_{\rm ext}L_{\rm d}/(4\pi cr_{\rm ext}^2)$, where $\xi_{\rm ext}$ is the covering factor of the medium reprocessing the accretion disk radiation, and `ext' stands either for `BLR' or `HDR'.

\begin{figure}
\includegraphics[width=\columnwidth]{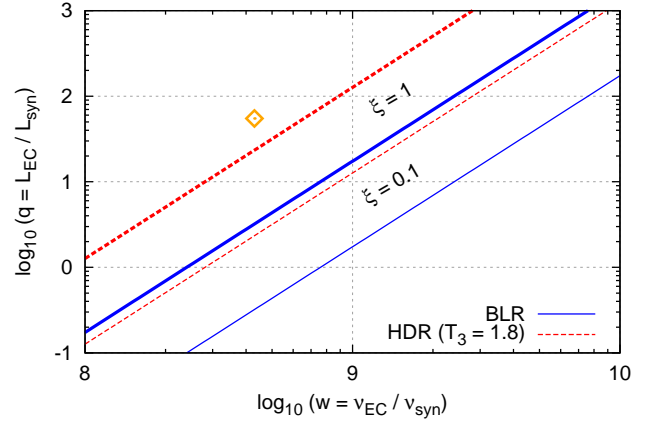}
\caption{Parameter space of the peak frequency ratio $w=\nu_{\rm EC}/\nu_{\rm syn}$ of the external-Compton to synchrotron components produced by electrons of the same energy, and the Compton dominance parameter $q=L_{\rm EC}/L_{\rm syn}$. For a given energy of external photons and a covering factor $\xi$ of the external radiation source, $q$ is a function of $w$. \emph{Solid lines}: constraints for the broad-line region; \emph{dashed lines}: constraints for the hot-dust region; \emph{thin lines}: $\xi=0.1$; \emph{thick lines}: $\xi=1$. \emph{The orange point} marks the location for PKS~1510$-$089 during the second \emph{Herschel} epoch (H2; MJD 55790).}
\label{fig_q_nu}
\end{figure}

As we argued in the previous section, the broad-band SED  up to the K band can be explained by a single synchrotron component. However, the GeV gamma-ray emission is most likely due to the Comptonization of external radiation (External-Compton; EC). Let us assume for a moment that these components are produced by the same population of ultra-relativistic electrons. This imposes two direct observational constraints. First, the luminosity ratio of the EC component to the synchrotron component, or the Compton dominance parameter, is $q=L_{\rm EC}/L_{\rm syn}\simeq 55$. Second, the frequency ratio of the peaks of the two components is $w=\nu_{\rm EC}/\nu_{\rm syn}\simeq 4.3\times 10^8$. \cite{2009ApJ...704...38S} showed that there is a direct relation between these two parameters that depends only on the covering factor $\xi_{\rm ext}$ and the energy of external photons in the external frame (see their equation 52). It can be expressed as:
\be
\xi_{\rm BLR} \simeq 0.6\times\frac{q_1}{w_9^2}
\,,
\quad
\xi_{\rm HDR} \simeq 1.7\times\frac{q_1}{w_9^2T_3^{5.2}}
\,,
\ee
where $T$ is the dust temperature.\footnote{These relations are valid as long as the EC process proceeds in the Thomson regime. The observed energy of EC radiation produced in the Thomson regime is $E_{\rm EC,obs} < (m_{\rm e}c^2)^2/[12(1+z)E_{\rm ext}] = 16\;{\rm GeV}/(E_{\rm ext}/1\;{\rm eV})$, where $E_{\rm ext}$ is the energy of external radiation in the external frame \citep{2010ApJ...721.1383A}. Even in the more constraining case of BLR with $E_{\rm ext} = 10\;{\rm eV}$, we find that $E_{\rm EC,obs} < 1.6\;{\rm GeV}$, which is satisfied in PKS~1510$-$089 by the observationally constrained spectral peak of the high-energy component.} These relations are shown in Figure \ref{fig_q_nu}, adopting $T_3=1.8$. Typically assumed values of the covering factor are $\xi_{\rm ext}\sim 0.1-0.3$. The observed values of $q$ and $w$ for PKS~1510$-$089 in the H2 state require $\xi_{\rm ext}\gg 1$, which is physically forbidden. Similar constraints on $q$ and $w$ also allow us to rule out the Synchrotron-Self-Compton (SSC) mechanism as the origin of the gamma-ray emission. Hence, \emph{it is not possible to fit the infrared and gamma-ray parts of the SED with a single-zone model.}

The observed synchrotron and EC components must be produced at distinct locations in the jet, where the local values of $q$ are different. Since the EC process proceeds in the Thomson regime, we have $q\simeq u_{\rm ext}'/u_{\rm B}'$, where $u_{\rm B}'=B'^2/(8\pi)$ is the magnetic energy density in the jet co-moving frame. We assume that the magnetic field scales as $B'\propto 1/r$, while the external radiation fields in the co-moving frame are approximated with $u'_{\rm ext}(r)\simeq (4/3)\Gamma_{\rm j}^2u_{\rm ext}/[1+(r/r_{\rm ext})^{\beta_{\rm ext}}]$, where $\Gamma_{\rm j}$ is the jet Lorentz factor, and we choose $\beta_{\rm BLR}=3$ and $\beta_{\rm HDR}=4$ \citep[see also][]{2012ApJ...754..114H}. We further assume that the observed gamma-ray emission is produced in the BLR -- this is supported by the variability time-scale of the order of days observed exclusively in the gamma-ray band. We model this component at $r_0=r_{\rm BLR}$, adopting $\Gamma_{\rm j}=20$, a half-opening angle $\theta_{\rm j}=1/\Gamma_{\rm j}$, and the covering factors $\xi_{\rm BLR}=\xi_{\rm HDR}=0.1$. The high Compton dominance is assured by taking a relatively weak magnetic field $B'=1\;{\rm G}\times (r_{\rm BLR}/r)$. We inject electrons with a broken power-law distribution of the random Lorentz factors, $N_\gamma\propto\gamma^{-p_{\rm i}}$, with $p_1=1.1$ for $\gamma<\gamma_{\rm br}$, and $p_2=4$ for $\gamma>\gamma_{\rm br}$. A very hard low-energy slope is necessary to avoid the EC component contributing to the observed X-ray emission. The injected electron energy distribution is softened by $\Delta p=1$ due to efficient cooling above a cooling break located at $\gamma_{\rm c}\simeq 15$. Our choice of $\gamma_{\rm br}=270$ places the EC peak in the low-energy end of the \emph{Fermi}/LAT range ($\sim 100\;{\rm MeV}$), and the synchrotron peak in the middle of the \emph{Herschel} range ($\sim 250\;{\rm\mu m}$). Parameters of this model are listed in Table \ref{tab_sed_models}.

\begin{table}
\centering
\caption{Parameters of SED models shown in Figure \ref{fig_sed}}
\label{tab_sed_models}
\begin{tabular}{cccc}
\hline\hline
  & H2 (BLR) & F2 (BLR) & H2 (HDR) \\
\hline
$r_0$ [pc]               & 0.07 & 0.07 & 3.2    \\
$\Gamma_{\rm j}$       & 20   & 20   & 20     \\
$B'(r)$ [G]            & 1    & 1    & 0.022  \\
$p_1$                  & 1.1  & 1.1  & 2.2    \\
$p_2$                  & 4    & 4    & 6      \\
$\gamma_{\rm min}$     & 1    & 1    & 1      \\
$\gamma_{\rm br}$      & 270  & 500  & $10^4$ \\
$\gamma_{\rm max}$     & $2\times 10^4$ & $2\times 10^4$ & $3\times 10^5$ \\
$K_{\rm e}$ [$s^{-1}$] & $3\times 10^{46}$ & $3\times 10^{46}$ & $3\times 10^{49}$ \\
\hline
\end{tabular}
\end{table}

The observed infrared emission must be produced in a region of low Compton dominance. Such a region cannot be found, at least for our parameter choice, between $r_{\rm BLR}$ and $r_{\rm HDR}$. Hence, we model this emission at $r_0=10^{19}\;{\rm cm} \simeq 5\,r_{\rm HDR}$. The injected electron energy distribution is a broken power-law with $p_1=2.2$, $p_2=6$ and $\gamma_{\rm br}=10^4$. The low-energy slope $p_1$ is chosen to match the \emph{Herschel} spectrum. The cooling is inefficient, and thus no cooling break is present. The EC component extends just below the observed X-ray emission and the gamma-ray emission in the 1--10 GeV range. On the low-energy end of the SED, we find the synchrotron emission to be self-absorbed below the frequency $\nu_{\rm abs}\sim 40\;{\rm GHz}$. This value is characteristic for the distance scale of a few pc \citep{2008ApJ...675...71S}, however, it is twice lower than the frequency of the spectral break detected by \emph{Planck} in February 2010 \citep{2011A&A...536A..15P}. The synchrotron self-absorption threshold frequency can be increased by allowing the jet to be more collimated, e.g. due to the formation of reconfinement shocks \citep{1997MNRAS.288..833K,2009MNRAS.392.1205N,2009ApJ...699.1274B}.

With a model of the H2 (low) spectral state on hand, we attempted to model a transition to the F2 (high) state. In Figure \ref{fig_sed}, we show a model of the F2 state obtained by varying a single parameter of the H2 state model, the break energy of the injected electron distribution of the SED component produced in the BLR \citep[see also][]{2009ApJ...705.1624A}, $\gamma_{\rm br}=500$ (instead of $\gamma_{\rm br}=270$). This resulted in a substantial increase of the gamma-ray flux and a modest increase of the IR flux, satisfying the observational constraint that the NIR, optical/UV and X-ray fluxes remain roughly constant. This scenario predicts a correlated variability between MIR and gamma-ray bands, i.e., that the F2 gamma-ray flare had a weak MIR counterpart. Unfortunately, we do not have simultaneous MIR data to verify this prediction. This simple model also underpredicts the gamma-ray flux at $1-2\;{\rm GeV}$, and thus an additional spectral component may be required in this energy range. While other scenarios of a spectral transition between H2 and F2 are certainly possible, this seems to be the only solution involving a change of a single parameter.

\section{Discussion}
\label{sec_dis}

The FIR spectrum of PKS~1510$-$089 measured by \emph{Herschel} is consistent with a simple power-law model. We did not find any direct evidence for a double synchrotron component, like the sharp spectral features observed by \cite{2011ApJS..195...19O} in 3C~454.3 and \cite{2012ApJ...754..114H} in 3C~279. However, indirect evidence suggests the existence of a second synchrotron component of much lower luminosity. We showed that it is not possible to fit the entire SED of PKS~1510$-$089 with a one-zone synchrotron-EC model, because the observed Compton dominance $q=L_{\rm EC}/L_{\rm SYN}$ is not compatible with the relative position of the synchrotron and EC peaks. The SED can be explained by a model consisting of two blazar zones characterized by different values of $q$, and thus spatially separated. In addition, the optical/UV spectrum suggests the presence of a thermal component, presumably produced in the accretion disk, with a further implication that the associated accretion disk corona can at least partly explain the X-ray emission.

This model is very tightly constrained by the observational data, and thus has several testable predictions. The thermal quasar emission is expected to be variable over very long time-scales (months/years), the component produced in the HDR should vary over weeks, and the component produced in the BLR over days. In our model, we should expect such variability time-scales in the optical/UV, infrared and gamma-ray bands, respectively. This is roughly consistent with the 2011 data for PKS~1510$-$089 presented in Figure \ref{fig_lc_mwl} and the long-term data shown in Figure \ref{fig_lc_long}. The fast optical flares observed in 2009 were significantly brighter and strongly polarized, and thus require a contribution of a synchrotron spectral component in the optical band. Such a component could extend to the FIR range, depending on the synchrotron self-absorption threshold. For a compact emitting region, typical for its location in the BLR, self-absorption could begin in the FIR range, producing a noticeable spectral break \citep{2012ApJ...754..114H}. For a large emitting region, typical for its location in the HDR, the self-absorption begins in the (sub-)mm range \citep{2008ApJ...675...71S}. Thus, in high gamma-ray/optical states, like the one observed in 2009, we expect two clear signatures in the FIR band of the synchrotron component produced in a compact region: a sharp spectral break; and variability on daily time-scales. Further FIR observations of this or other luminous blazars are necessary to test these predictions.

Our model is different from that of \cite{2010ApJ...721.1425A}, who analyzed the 2009 active state of PKS~1510$-$089. We first note that they adopted different electron energy distributions: with $p\simeq 3.2$ and $\gamma_{\rm max}=2.2\times 10^4$, their synchrotron components were relatively high and soft, extending into the far-UV band (they adopted $B'=1\;{\rm G}$ and $\Gamma_{\rm j}\simeq 15$). Unbeknownst to these authors, their synchrotron models are rather consistent with our \emph{Herschel} data, at least in the March 2009 state, thanks to the introduction of a break in the electron energy distribution at $\gamma_{\rm br}\simeq 200$. However, these models are not consistent with very soft NIR spectra that we identify in the SMARTS data, and that, to a lesser degree, can be seen in Figure 24 of \cite{2010ApJ...721.1425A}. As we show in Figure \ref{fig_colors}, the $K-J$ color in the H2 epoch is quite typical for this source. 

To explain the 2009 flaring state, when the optical/NIR flux was well correlated with the gamma-ray flux, in our two-zone model, one of the synchrotron components should dominate the thermal accretion disk emission. The fast variability of the 2009 flares indicates that it should be the component produced at shorter distance scale within the jet, i.e., the one located in the BLR \citep{2010MNRAS.405L..94T}. Now, we know that the value of Compton dominance varied in the range of $q\sim 10$ --- $50$. Our BLR component has a very large $q\simeq 200$ due to a rather low local magnetic field strength. Hence, the 2009 activity could have been accompanied by a significant increase of the magnetic field, which can be achieved via compression by a strong shock wave. Indeed, \cite{2010ApJ...710L.126M} report a superluminal knot observed with VLBA at $7\;{\rm mm}$, the emergence of which (passage through the $7\;{\rm mm}$ radio core) roughly coincided with the main gamma-ray/optical flare. Also during that flare, a strong increase in the optical polarization degree was observed \citep[see also][]{2011PASJ...63..489S}, which is consistent with a strong shock wave compressing the magnetic fields. Hence, the 2009 activity of PKS~1510$-$089 was most likely caused by additional dissipation provided by a passing shock wave, and apparently in the summer of 2011 such an additional factor was not present.

\cite{2008ApJ...672..787K} and \cite{2010ApJ...716..835A} observed PKS~1510$-$089 in 2006 and 2009 with the \emph{Suzaku} X-ray telescope and various other facilities. The focus of their work was on the soft X-ray part of the SED, but they also measured a hard optical/UV spectrum, which they interpreted as thermal emission from the accretion disk. They adopted a soft synchrotron component peaking in the FIR range. Using non-simultaneous data, they noticed a very soft NIR spectrum and interpreted it as an excess resulting from the starlight of the host galaxy. The long-term SMARTS data invalidate this interpretation, because they show that the large-amplitude NIR variability is not associated with a correlated variability of the $K-J$ color. The strong variability amplitude in the NIR band can be explained only by the synchrotron emission. Moreover, a hard synchrotron component inferred from our \emph{Herschel} observations is consistent with previous observations of PKS~1510$-$089 by the \emph{Planck} and \emph{Spitzer} satellites (\citealt{2011A&A...536A..15P,2011ApJ...732..116M}; see Figure \ref{fig_sed}). A similar spectral shape of the synchrotron component in PKS 1510$-$089 was adopted by \cite{2009A&A...508..181D}.

The possibility that the X-ray emission of PKS~1510$-$089 is produced at least partly in the accretion disk corona was considered neither by \cite{2008ApJ...672..787K} nor by \cite{2010ApJ...716..835A}, even though the X-ray flux measured with \emph{Suzaku} is comparable to that presented in this work. The long-term X-ray light curves of PKS~1510$-$089 presented by \cite{2010ApJ...710L.126M} indicate a flux lower limit of $5\times 10^{-12}\;{\rm erg\,s^{-1}\,cm^{-2}}$ in the $2.4$ -- $10\;{\rm keV}$ range, which corresponds to $8\times 10^{-12}\;{\rm erg\,s^{-1}\,cm^{-2}}$ in the $0.3$ -- $10\;{\rm keV}$ range for a photon index of $\Gamma_{\rm X}=1.5$. Our estimate of the $0.3$ -- $10\;{\rm keV}$ X-ray flux attributed to the accretion disk corona is $\sim 6\times 10^{-12}\;{\rm erg\,s^{-1}\,cm^{-2}}$, which is consistent with the lower limit given above. Another possible signature of the coronal emission contributing to the X-ray band is fluorescent Fe emission line. However, even the very deep \emph{Suzaku} observations reported by \cite{2008ApJ...672..787K} do not reveal any hint of such lines, although, such lines are generally hard to detect even in intrinsically similar sources with misaligned jets (Steep-Spectrum Radio Quasars and Broad-Line Radio Galaxies; e.g. \citealt{2006ApJ...642..113G,2011ApJ...727...19F}). We also note that our \emph{Swift}/XRT data are of insufficient quality to verify the presence of the soft X-ray excess detected by \emph{Suzaku}.

Our inference of two separate energy dissipation regions (`blazar zones') in AGN jets is consistent with the works of \cite{2011ApJS..195...19O} and \cite{2012ApJ...754..114H}. If confirmed by further comprehensive studies of multiwavelength emission of blazars, it has significant implications for the long-standing theoretical problem of the location of blazar zones and the underlying mechanisms of energy dissipation and particle acceleration. The answer to this puzzle may turn out to be quite complex. At the distance scale of $\sim 0.1\;{\rm pc}$, in the BLR, possible dissipation mechanisms could be internal shocks, produced by collisions of jet portions of high Lorentz factor contrast \citep{1994ApJ...421..153S,2001MNRAS.325.1559S,2010MNRAS.405L..94T}, or magnetic reconnection enabled by global magnetic field reversals \citep{2011MNRAS.413..333N} or current-driven instabilities (\citealt{2006A&A...450..887G}; Nalewajko \& Begelman, submitted). At the distance scale of $3\;{\rm pc}$, in the HDR, dissipation could proceed via reconfinement shocks, produced by interaction of the jet with the external medium \citep{1988ApJ...334..539D,2008ApJ...675...71S,2012MNRAS.420L..48N}, and possibly driving turbulence \citep{2012arXiv1201.5402M}. The need for distinct particle acceleration mechanisms is underlined by the different energy distributions of injected electrons required to explain the observational data. A hard low-energy electron index for the component produced in the BLR, $p_1=1.1$, suggests magnetic reconnection \citep[\eg,][]{2001ApJ...562L..63Z,2008ApJ...682.1436L}, while the one for the component produced in the HDR, $p_1=2.2$, constrained directly by the \emph{Herschel} data, may favor the shock acceleration \citep[\eg,][]{1998PhRvL..80.3911B}. Thus, a possible scenario for the overall activity of PKS~1510$-$089 may involve dissipation via magnetic reconnection at sub-pc scales and additional dissipation via recollimation shocks at supra-pc scales. Strong breaks in the injected electron energy distributions, with $p_2-p_1\simeq 3-4$, may indicate the variation of $\gamma_{\rm max}$ along the propagation of the emitting region. That $\gamma_{\rm br}$ is much larger in the HDR models than in the BLR models is consistent with less efficient cooling and/or longer source evolution time scale in the HDR. However, a definite theory of particle acceleration in relativistic sources is necessary to explain the observed spectral breaks.

\section{Conclusions}
\label{sec_con}

We observed blazar PKS~1510$-$089 with the \emph{Herschel} Space Observatory, using its PACS and SPIRE photometric instruments. We detected the source consistently with all 6 filters at 5 epochs in the relatively quiet state from mid July to early September 2011. We did not find a significant variability amplitude in the FIR range. The FIR SED for each epoch is consistent with a power-law model, with a slight harder-when-brighter trend.

We collected simultaneous multiwavelength data from \emph{Fermi}/LAT, \emph{Swift}, SMARTS and SMA, to place our \emph{Herschel} observations within a broader context. Analysis of the short-term multiwavelength light curves indicates a low fractional variability in all bands between the millimeter and X-ray, accompanied by two gamma-ray flares directly preceding the \emph{Herschel} observations. Broad-band SEDs were extracted for two epochs -- the second \emph{Herschel} epoch (`H2') and the second gamma-ray flare (`F2'). They show different gamma-ray spectra, with the flaring state spectrum being harder and more complex than the quiet state spectrum. They also show a consistent spectral structure in the NIR/optical/UV range -- a very soft NIR ($K-J$) spectrum and a hard optical/UV spectrum. We also compare the long-term gamma-ray and optical/NIR activities, using the \emph{Fermi}/LAT and SMARTS data. The SMARTS data reveal the existence of a lower limit on flux in $J$ and $R$ filters, and a noticeably different behavior of the $K$ flux. The $K-J$ color does not depend on the $J$ luminosity, in contrast to the `optical-$J$' colors, which show the typical `redder-when-brighter' trends.

We interpret the optical/UV spectrum in terms of  thermal emission from the accretion disk. This is supported by the hard spectrum and the existence of the lower limit to the flux. The associated accretion disk corona can partly explain the X-ray spectrum. The soft NIR spectrum is interpreted as the high-energy cut-off in a synchrotron component. This component cannot be produced in the same region as the main gamma-ray emission for two reasons: 1) their variations are not correlated, 2) in the synchrotron-EC scenario using a single population of electrons, the relation between the Compton dominance parameter $q=L_{\rm EC}/L_{\rm syn}\simeq u_{\rm ext}'/u_{\rm B}'$ and the emitted frequency ratio $w=\nu_{\rm EC}/\nu_{\rm syn}$ is strongly constrained. A one-zone leptonic model would require an unrealistically high energy density of the external radiation to match the NIR and gamma-ray spectra simultaneously. We consider a two-zone model, with the infrared emission produced in the jet region of a small $q$ and the gamma-ray emission produced in the region of a very large $q$. We find a consistent model, in which the high-$q$ region is associated with the broad-line region, and the low-$q$ region is located in the hot-dust region. We show that `orphan' gamma-ray flares can be explained by varying solely the break energy of the electron energy distribution injected in the high-$q$ (BLR) region. Hence, we identify the \emph{Herschel} results mainly with the synchrotron emission produced at the supra-pc scale, and the two gamma-ray flares with the EC (BLR) component produced at the sub-pc scale.

\acknowledgments

We acknowledge financial support by the Polish NCN grant DEC-2011/01/B/ST9/04845, the NSF grant AST-0907872, the NASA ATP grant NNX09AG02G, the NASA Fermi GI grant NNX11AO39G, the NASA JPL/IPAC Herschel GO grant RSA~1433865 to Stanford University, and a DoE support to SLAC via contract DE-AE3-76SF00515. We acknowledge support from the Faculty of European Space Astronomy Center (ESAC).

\emph{Herschel} is an ESA space observatory with science instruments provided by European-led Principal Investigator consortia and with important participation from NASA.

The \textit{Fermi}-LAT Collaboration acknowledges generous ongoing support from a number of agencies and institutes that have supported both the development and the operation of the LAT as well as scientific data analysis. These include the National Aeronautics and Space Administration and the Department of Energy in the United States, the Commissariat \`a l'Energie Atomique and the Centre National de la Recherche Scientifique / Institut National de Physique Nucl\'eaire et de Physique des Particules in France, the Agenzia Spaziale Italiana and the Istituto Nazionale di Fisica Nucleare in Italy, the Ministry of Education, Culture, Sports, Science and Technology (MEXT), High Energy Accelerator Research Organization (KEK) and Japan Aerospace Exploration Agency (JAXA) in Japan, and the K.~A.~Wallenberg Foundation, the Swedish Research Council and the Swedish National Space Board in Sweden. Additional support for science analysis during the operations phase is gratefully acknowledged from the Istituto Nazionale di Astrofisica in Italy and the Centre National d'\'Etudes Spatiales in France.

We acknowledge the use of NASA's \emph{Swift} Observatory.

The Submillimeter Array (SMA) is a joint project between the Smithsonian Astrophysical Observatory and the Academia Sinica Institute of Astronomy and Astrophysics and is funded by the Smithsonian Institution and the Academia Sinica.

\end{document}